\documentclass[conference]{IEEEtran}
\IEEEoverridecommandlockouts
\usepackage{cite}
\usepackage{amsmath,amssymb,amsfonts}
\usepackage{algorithmic}
\usepackage{graphicx}
\usepackage{textcomp}
\usepackage{xcolor}

\usepackage{times}
\usepackage{epsfig}
\usepackage{amsmath}
\usepackage{amssymb}
\usepackage{subcaption}
\usepackage{booktabs}
\usepackage{float} 
\usepackage{color}
\usepackage{cite}
\usepackage{enumitem}

\def\BibTeX{{\rm B\kern-.05em{\sc i\kern-.025em b}\kern-.08em
    T\kern-.1667em\lower.7ex\hbox{E}\kern-.125emX}}
\begin{document}

\title{MGFI-Net: A Multi-Grained Feature Integration Network for Enhanced Medical Image Segmentation}

\author{\IEEEauthorblockN{1\textsuperscript{st} Yucheng Zeng}
\IEEEauthorblockA{\textit{dept. Shanghai University} \\
Shanghai, China \\
544722646a@gmail.com}
}

\maketitle

\begin{abstract}

Medical image segmentation plays a crucial role in various clinical applications. A major challenge in medical image segmentation is achieving accurate delineation of regions of interest in the presence of noise, low contrast, or complex anatomical structures. Existing segmentation models often neglect the integration of multi-grained information and fail to preserve edge details, which are critical for precise segmentation. To address these challenges, we propose a novel image semantic segmentation model called the Multi-Grained Feature Integration Network (MGFI-Net). Our MGFI-Net is designed with two dedicated modules to tackle these issues. First, to enhance segmentation accuracy, we introduce a Multi-Grained Feature Extraction Module, which leverages hierarchical relationships between different feature scales to selectively focus on the most relevant information. Second, to preserve edge details, we incorporate an Edge Enhancement Module that effectively retains and integrates boundary information to refine segmentation results. Extensive experiments demonstrate that MGFI-Net not only outperforms state-of-the-art methods in terms of segmentation accuracy but also achieves superior time efficiency, establishing it as a leading solution for real-time medical image segmentation.
\end{abstract}


\section{Introduction}
Medical Image Segmentation (MIS) aims to classify each pixel in a medical image, dividing it into distinct regions such as polyp\cite{polyp,Ali2023,Dumitru2023}, nuclei\cite{bowl,10.1007/978-3-319-24574-4_28,Lee2022}, or lesions\cite{isic,Tschandl2018}. This process is crucial in medical image analysis and has numerous practical applications\cite{Cheng2013,Fu2018,blood,5306149} in clinical fields. Given its importance, MIS has been extensively researched, leading to many notable advancements in the field. However, as a specialized area of MIS, semantic segmentation in complex medical images, such as those with noisy backgrounds\cite{polyp}, delicate nuclear boundaries\cite{bowl}, or ambiguous boundaries\cite{isic}, remains a persistent challenge. 


Many existing models for medical image segmentation primarily focus on extracting local features to delineate regions of interest\cite{unet,u1+,u2+}. These models often achieve reasonable performance in standard conditions\cite{ZHANG2023109020}, but they struggle when applied to medical images with low contrast\cite{YUAN2024110428}, complex structures\cite{YUAN2019248}, or noise interference\cite{polyp}. One common limitation is their inability to effectively combine detailed local information with broader contextual cues. Without this multi-grained\cite{Yan_2017_ICCV} feature integration, these models often fail to capture the full complexity of the image, leading to incomplete or inaccurate segmentation. To address the challenge of boundary preservation, many segmentation models incorporate edge detection techniques\cite{ce,u3+}. While these approaches help to refine object boundaries, they often fall short when applied to segmentation targets with complex or irregular morphologies. This is primarily because traditional edge detection mechanisms rely on fixed receptive fields and are unable to adaptively model the geometric variations of objects. As a result, the boundaries in regions with intricate shapes or surrounding noise often remain blurred or poorly defined.

\begin{figure}[t]
    \centering
    \begin{subfigure}[b]{0.15\textwidth}  
        \centering
        \includegraphics[width=0.8\textwidth]{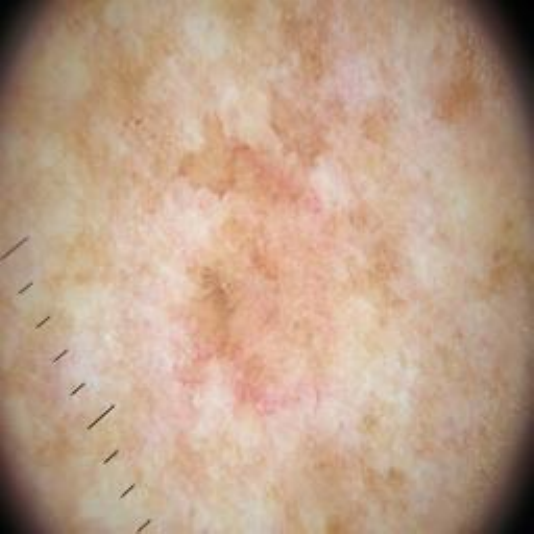}  
        \caption{}
    \end{subfigure}
    \begin{subfigure}[b]{0.15\textwidth}  
        \centering
        \includegraphics[width=0.8\textwidth]{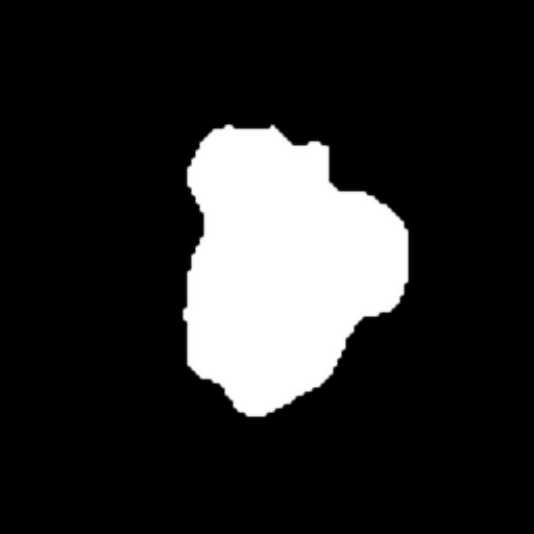}  
        \caption{}
    \end{subfigure}
    \begin{subfigure}[b]{0.15\textwidth}  
        \centering
        \includegraphics[width=0.8\textwidth]{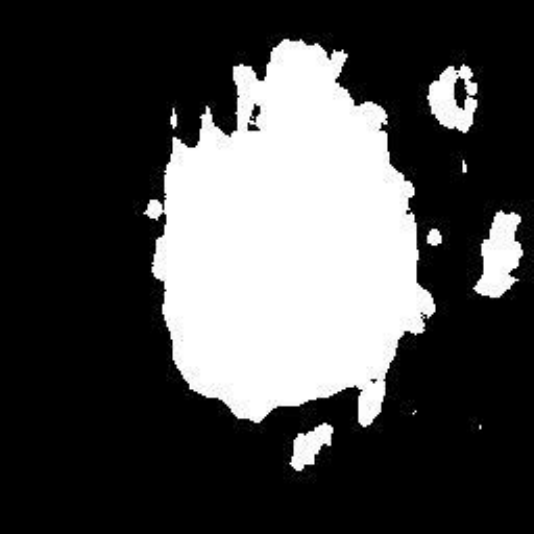}  
        \caption{}
    \end{subfigure}
    \begin{subfigure}[b]{0.15\textwidth}  
        \centering
        \includegraphics[width=0.8\textwidth]{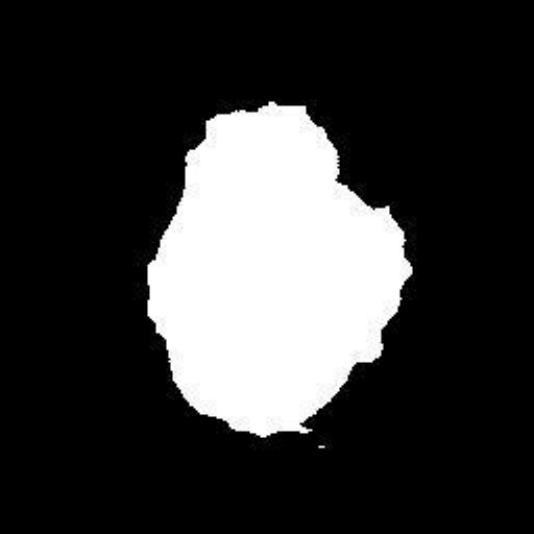}  
        \caption{}
    \end{subfigure}
    \caption{This set of images shows segmentation results before and after applying the Adaptive Edge module. (a) The original image, (b) the ground truth, (c) the result without the Adaptive Edge module, with blurred edges, and (d) the result with the module, showing better edge preservation and accuracy. This figure demonstrates the effectiveness of the module in capturing complex edge details and improving segmentation performance.}
    \label{fig2}
\end{figure}

To address these challenges, we propose a  image semantic segmentation model called the Multi-Grained Feature Integration Network (MGFI-Net). This model integrates multi-grained features, combining both local details and broader contextual information to improve segmentation accuracy, especially when dealing with complex morphologies and noisy images. The key innovation of our model lies in its ability to capture hierarchical relationships across different feature scales\cite{10.1109/TCSVT.2023.3317486}, allowing it to focus selectively on internal structures and important boundaries. 
However, while this multi-grained approach significantly enhances performance, it can sometimes result in the loss of fine edge details\cite{Chen_2018_ECCV}. To overcome these limitations, we introduce an Adaptive Edge (AE) module, which incorporates deformable convolutions\cite{8237351,8953797,10203769,10655307} to dynamically adjust the shape and position of convolutional kernels, enabling the model to better capture geometric variations of segmentation targets. Unlike traditional edge detection methods, the AE module refines edge information by adaptively preserving the clarity of boundaries, even in regions with complex morphologies or surrounding noise. As shown in Fig.~\ref{fig2}, this module effectively integrates contextual and edge-specific features, improving both segmentation accuracy and robustness. By leveraging the AE module alongside multi-grained feature integration, MGFI-Net addresses the challenges of integrating multi-grained contextual information and edge blurring, making it well-suited for medical image segmentation tasks.

In summary, the contributions of this paper are as follows:
\begin{enumerate}[label=\arabic{enumi})]
\item To address the challenge of integrating multi-grained information for better segmentation, we propose MGFI-Net. The model leverages multi-grained features to handle complex backgrounds, noise, and low-contrast images, which significantly improves segmentation accuracy.
\item To extract both multi-grained features and complex edge information, we propose two modules: the Multi-Grained Feature Integration (MGFI) module and the Adaptive Edge (AE) module. The MGFI module is designed to capture hierarchical relationships between feature scales, enabling the model to focus on the most relevant information for segmentation. The AE module, using deformable convolution, is designed to dynamically adjust convolutional kernels to adaptively learn and preserve intricate edge details. Together, these modules enhance both segmentation accuracy and edge preservation, improving the model’s ability to handle complex morphology and noise interference.
\item We evaluate the performance of MGFI-Net on three public medical image segmentation datasets, including polyp, nucleus, and skin lesion segmentation. The results show that MGFI-Net achieves superior segmentation accuracy compared to other state-of-the-art (SOTA) models across multiple metrics. Additionally, we evaluate time efficiency using metrics such as FLOPs, Params and FPS, demonstrating that MGFI-Net balances precise segmentation with computational efficiency. This highlights the capability of MGFI-Net to deliver accurate and efficient segmentation in challenging medical imaging tasks.
\end{enumerate}

\section{Related Work}
In recent years, deep learning has rapidly advanced, particularly with the development of Convolutional Neural Networks (CNN) and their variants, significantly improving medical image segmentation task\cite{ZHANG2023109020}. 

 One of the influential models in this domain is U-Net\cite{unet}, which uses skip connections to preserve detailed information during downsampling, making it highly effective for medical image segmentation. However, U-Net primarily focuses on local features and struggles with capturing broader contextual information, limiting its performance in images with complex structures. Several models have been proposed to improve both edge detection and contextual understanding in medical image segmentation. Attention U-Net\cite{atunet}, introduced to refine U-Net, incorporates an attention mechanism that focuses on relevant regions such as organ boundaries or lesions, improving segmentation precision. However, it still faces challenges in preserving edge details while capturing broader contextual information. CE-Net\cite{ce} improves multi-scale feature extraction by using dilated convolutions and a context encoder, making it suitable for handling complex anatomical structures. Despite this, CE-Net faces challenges with accurately segmenting unclear or blurred boundaries. KiU-Net\cite{DBLP:journals/tmi/ValanarasuSHP22} tackles edge preservation by using an overcomplete convolutional structure that enhances detail retention, particularly in edge-aware tasks. However, this design comes at the cost of higher computational complexity, making it less efficient for large-scale datasets.

\section{Method}
\begin{figure*}[t]
\begin{center}
\includegraphics[width=0.8\linewidth]{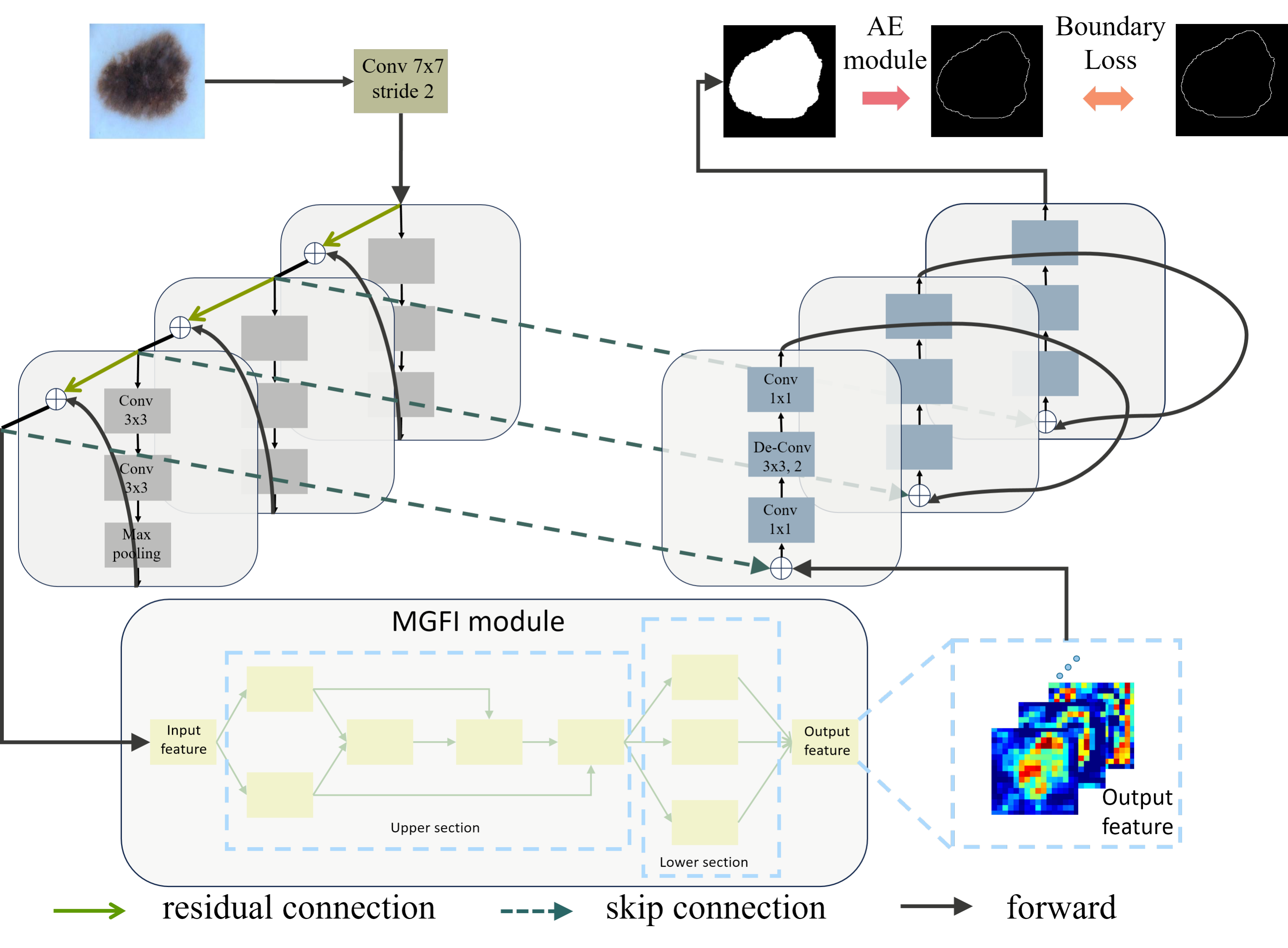}
\end{center}
   \caption{The architecture of the proposed MGFI-Net for medical image segmentation. The model starts with a convolutional encoder that extracts hierarchical features, which are then processed through the MGFI module. This module  selectively focuses on the most relevant information for accurate region delineation. Meanwhile, the AE module, which applies deformable convolutions, improves edge segmentation by refining boundary details.}
\label{fig3}
\end{figure*}

We propose the MGFI-Net for medical image segmentation, which consists of an encoder, a MGFI module, a decoder, and an Adaptive Edge (AE) module. The encoder extracts multi-scale information, while the decoder restores resolution, ensuring both high-level semantics and fine-grained details are preserved. The core of the model is the MGFI module, which leverages multi-grained information through multi-branch convolutions to capture detailed spatial information and hierarchical contextual relationships. To make edge details more accurate, we introduce an AE module at the end of the network, which employs deformable convolutions to adaptively refine complex edge structures, improving segmentation accuracy, particularly for challenging medical images with unclear boundaries. The complete network structure is illustrated in Fig.~\ref{fig3}.

\subsection{Multi-Grained Feature Integration module}
\begin{figure*}[t]
\begin{center}
\includegraphics[width=0.8\linewidth]{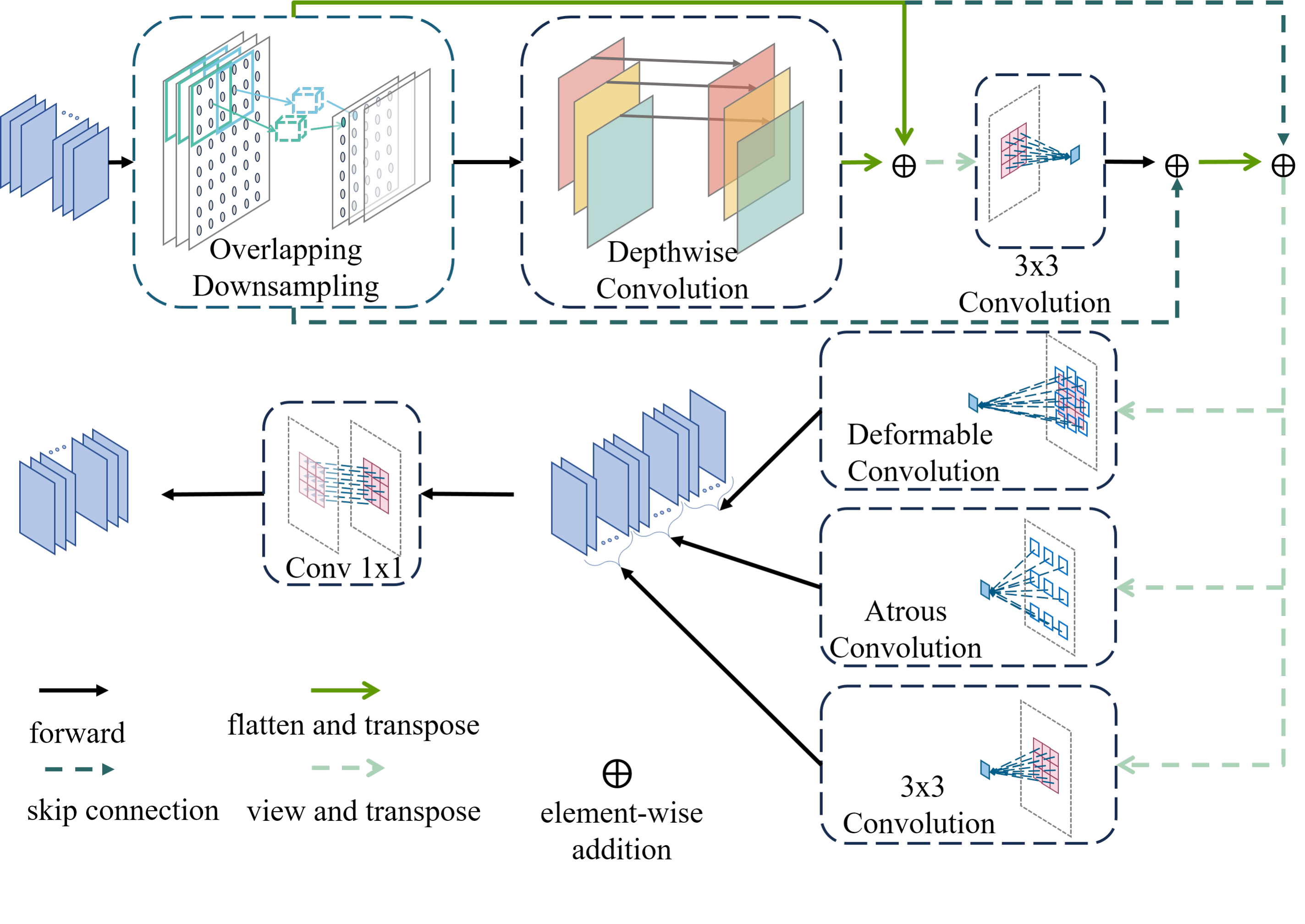}
\end{center}
   \caption{The structure of the MGFI module. The upper part of the module is responsible for extracting local features and global context information. The lower part includes three different convolution types: deformable convolution, atrous convolution, and standard convolution. These are applied in parallel to capture multi-grained features, which are then fused through channel concatenation.}
\label{fig4}
\end{figure*}

The MGFI module, as shown in Fig.~\ref{fig4}, is designed to address the limitations in capturing complex contextual information at different granularities, which can affect segmentation accuracy. Fortunately, the encoder extracts multi-scale features, and we further enhance this by refining and integrating both finer details and broader contextual relationships through the MGFI module. This integration across multiple granularity levels improves both detail preservation and contextual understanding, ultimately boosting the overall performance of the model.

To increase the ability of the model to capture diverse feature information, the MGFI module incorporates two main sections. The upper section focuses on extracting local features from the output of the encoder, while the lower section utilizes a multi-branch structure to further refine the features and integrate the previously extracted overlapping information. In the upper section, the input feature map $F_{in}$ undergoes overlapping downsampling, which reduces spatial dimensions while preserving continuity between regions. This downsampling results in a feature map, $F_{overlap}$, which is then processed by a local feature extraction branch. To further refine the extracted local features and enhance their representation, depthwise separable convolutions\cite{Chen_2018_ECCV} are applied, resulting in a refined feature map denoted as $F_{dw}$. The depthwise convolution operation can be mathematically represented by the \eqref{eq4}: 
\begin{equation}
    \label{eq4}
    y_{\text{depthwise}}^{(m)}(p_0) = \sum_{k=1}^{K} w_k^{(m)} \cdot x^{(m)}(p_0 + p_k)
\end{equation}

In this equation $y_{depthwise}^{(m)}(p_{0})$ represents the output of the depthwise convolution for the $m$-th channel, where $w_{k}^{(m)}$ are the filter weights, and $x^{(m)}(p_{0}+p_{k})$ are the input feature values at different positions. These depthwise convolutions efficiently capture local details while reducing the number of parameters. The following part then refines the local features and integrates the ${F_{overlap}}$ information back into the model, ensuring that no crucial spatial details are lost during the downsampling process, thereby improving the model's ability to retain both fine details and broader contextual information.

Once the local features are extracted and the overlapping information is re-integrated, they are fused along the channel dimension and refined through a 3×3 convolution, followed by batch normalization (BatchNorm) and a ReLU activation function. This step ensures that the fine-grained local details and the previously preserved overlapping spatial information are effectively combined. The feature map $F_{dw}$, produced by the depthwise separable convolution, is fused with the overlapping feature map $F_{overlap}$ to capture both local details and the spatial continuity across regions. Finally, this fusion and refinement process generates the feature map $F_{concat}$.The fusion process is mathematically represented in \eqref{eq3}:

\begin{equation}
    \label{eq3}
    F_{\text{concat}} = \text{ReLU} \left( \text{BatchNorm} \left( \text{Conv3}(F_{\text{overlap}} + F_{\text{dw}}) \right) \right)
\end{equation}

A residual connection mechanism is then employed to integrate the features generated by both branches. After overlapping downsampling, the input feature map $F_{in}$ produces a local feature representation, which then undergoes depthwise separable convolution to efficiently capture local details. The resulting feature map is further refined through a 3×3 convolution, batch normalization, and a ReLU activation function. The output from this process, which is the feature map $F_{concat}$, is then element-wise added to the overlapping feature map, $F_{overlap}$. Following this, both the refined feature map and $F_{overlap}$ are flattened, transposed, and then added together element-wise. This residual connection process helps preserve spatial details during the downsampling and convolution processes, preserving the continuity of features across regions. By fusing the features at multiple stages, the module enhances the robustness and learning capacity of the network, particularly when handling complex features. This approach maintains feature fidelity, improves overall feature representation, and prevents gradient vanishing, thus accelerating training convergence.

In the lower section, the MGFI module applies a multi-branch convolutional structure to refine the fused feature maps. To dynamically adjust the sampling locations of the convolution kernel and handle variations in target shapes and non-rigid deformations, the first branch is used with deformable convolutions, making it especially useful for irregularly shaped medical images. However, deformable convolutions alone may not capture broader contextual information, so the second branch is introduced with atrous convolutions, which expand the receptive field and capture a wider range of context without additional computational costs. While atrous convolutions provide this broader perspective, they may miss finer local details, so the third branch is used with standard 3×3 convolutions to ensure that important spatial details are preserved during the processing of broader features.

After processing the feature maps through the three branches, the outputs are concatenated along the channel dimension, and a 1×1 convolution is applied to compress and refine the fused features. This produces the final output feature map, $F_{final}$, which integrates information from various granularities, Improving the ability of the model to adapt to complex structures and shapes in medical images.

The feature maps from all three branches—deformable, atrous, and standard convolutions—are concatenated and combined to form the final output feature map $F_{final}$, as expressed in \eqref{eq2}:
\begin{equation}
    \label{eq2}
     F_{\text{final}} = \text{CONV} \left( \text{Concat}(F_{\text{deform}}, F_{\text{atrous}}, F_{\text{standard}}) \right)
\end{equation}

Here, Concat represents the concatenation operation. CONV refers to the 1×1 convolution that is applied to the concatenated feature maps to compress and refine the information. 

Overall, the MGFI module plays a critical role in addressing the limitations of the encoder by using specialized branches to refine and integrate local and global features. This process of multi-grained feature integration enhances the overall segmentation accuracy and adaptability of the model, particularly in challenging medical imaging tasks involving complex structures and noisy backgrounds.

\subsection{Adaptive Edge module}
In medical image segmentation tasks, accurately capturing edge information is crucial for high segmentation performance. However, preserving fine edge details, particularly in regions with complex or irregular boundaries, remains a challenge. To address this issue, we introduce the Adaptive Edge (AE) module, which enhances edge preservation by dynamically refining boundary information, ensuring precise segmentation in challenging regions.

The AE module utilizes deformable convolution, which differs from standard convolution by dynamically adjusting the sampling locations of the convolutional kernel through calculated offsets. This allows the model to better capture edges in regions with irregular shapes or significant deformation. First, the input feature map undergoes a 3×3 convolution operation to generate offsets. These offsets are output across multiple channels, determining the sampling positions for the deformable convolutional kernels.

Next, the deformable convolution uses the generated offsets to dynamically adjust its sampling locations based on the input feature map, enabling the model to better perceive and capture detailed edge information. This process helps the model to adaptively respond to complex or irregular boundaries, improving edge clarity in the segmented output. Finally, to simplify the edge information, the multi-channel output from the deformable convolution is processed through a 1×1 convolution, compressing it into a single-channel edge feature map. This refined edge map is then used for further edge supervision within the network, ensuring that the model accurately retains and utilizes edge details in its final segmentation predictions.

The deformable convolution is mathematically represented as \eqref{eq1}:
\begin{equation}
    \label{eq1}
     y(p_0) = \sum_{k=1}^{K} w_k \cdot x(p_0 + p_k + \Delta p_k)
\end{equation}

In this equation, $y(p_{0})$ represents the output at position $p_{0}$, $w_{k}$ is the weight at position $k$, $x(p_{0}+p_{k}+{\Delta}p_{k})$ is the input feature at the dynamically adjusted position $p_{0}+p_{k}+{\Delta}p_{k}$, and ${\Delta}p_{k}$ represents the learned offset. These offsets enable the convolution to adapt its sampling locations based on the structure of the input features.

By incorporating the AE module, MGFI-Net effectively preserves crucial edge information, enhancing segmentation accuracy in medical images, particularly in regions with complex structures or irregular boundaries.

\begin{figure*}[t]
    \centering
    \begin{subfigure}[b]{0.15\textwidth}  
        \centering
        \includegraphics[width=0.8\textwidth]{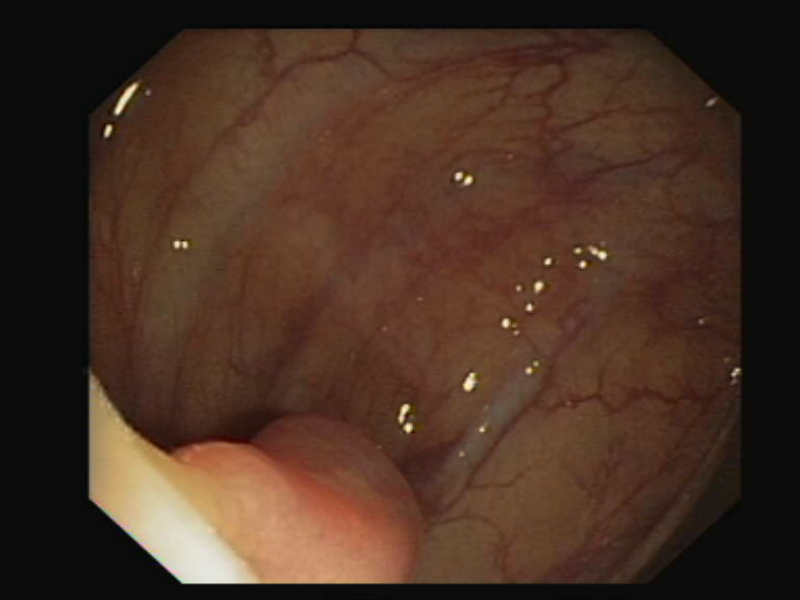} 
        \caption*{Input}
    \end{subfigure}
    \begin{subfigure}[b]{0.15\textwidth}  
        \centering
        \includegraphics[width=0.8\textwidth]{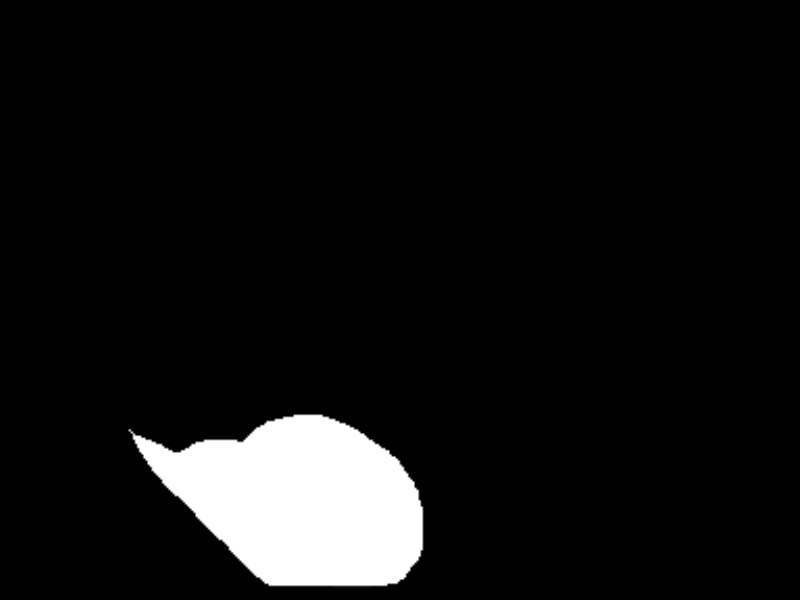} 
        \caption*{Ground Truth}
    \end{subfigure}
    \begin{subfigure}[b]{0.15\textwidth}  
        \centering
        \includegraphics[width=0.8\textwidth]{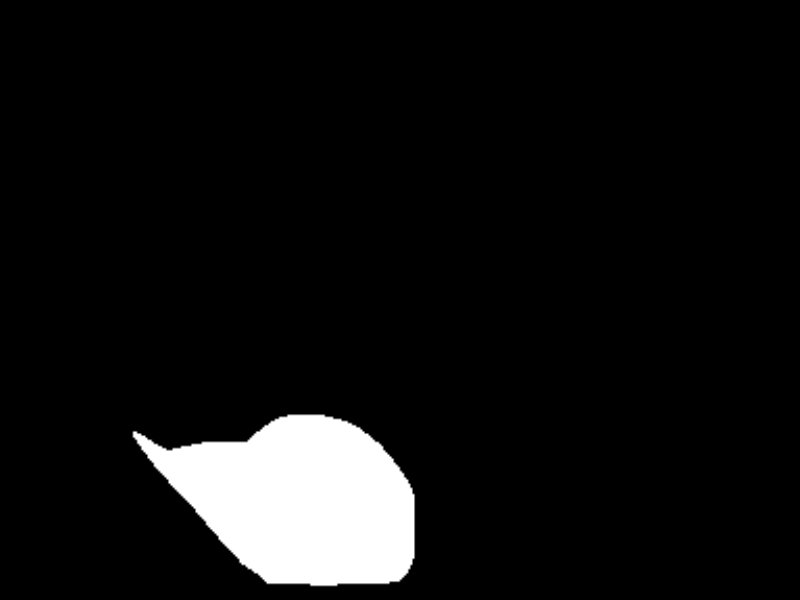}  
        \caption*{MGFI-Net}
    \end{subfigure}
    \begin{subfigure}[b]{0.15\textwidth}  
        \centering
        \includegraphics[width=0.8\textwidth]{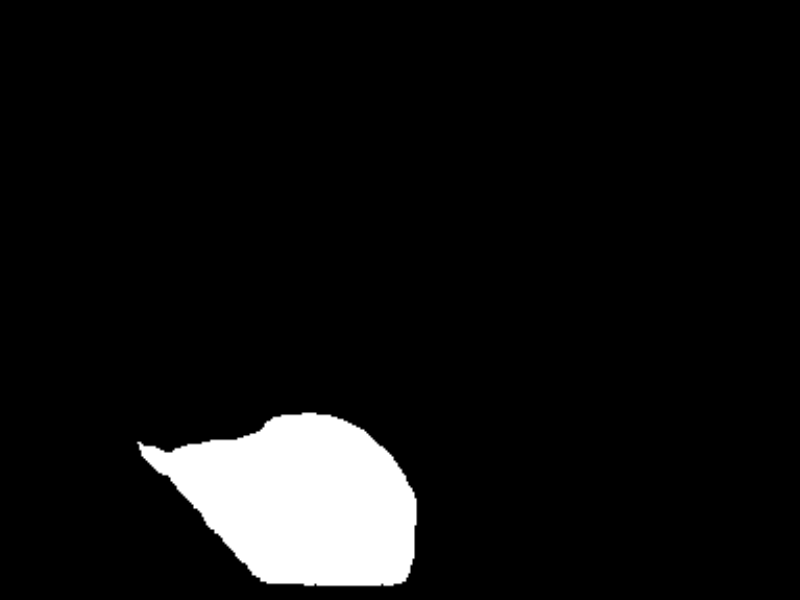}  
        \caption*{U-Net}
    \end{subfigure}
    \begin{subfigure}[b]{0.15\textwidth} 
        \centering
        \includegraphics[width=0.8\textwidth]{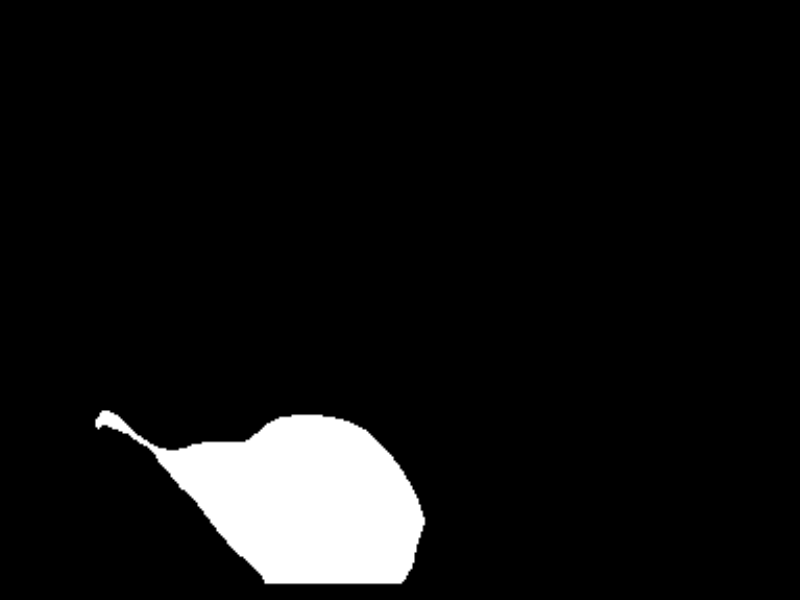}  
        \caption*{UNet++}
    \end{subfigure}
    \begin{subfigure}[b]{0.15\textwidth}  
        \centering
        \includegraphics[width=0.8\textwidth]{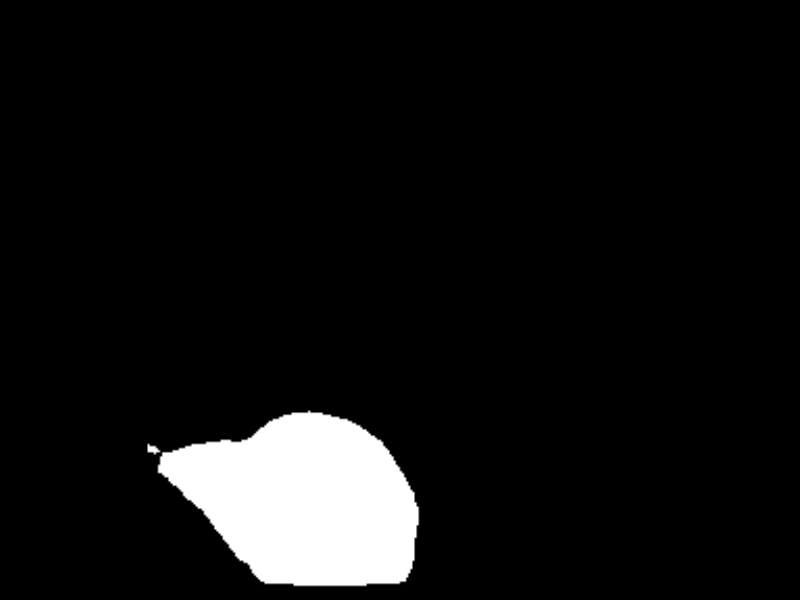}  
        \caption*{CE-Net}
    \end{subfigure}
    \begin{subfigure}[b]{0.15\textwidth} 
        \centering
        \includegraphics[width=0.8\textwidth]{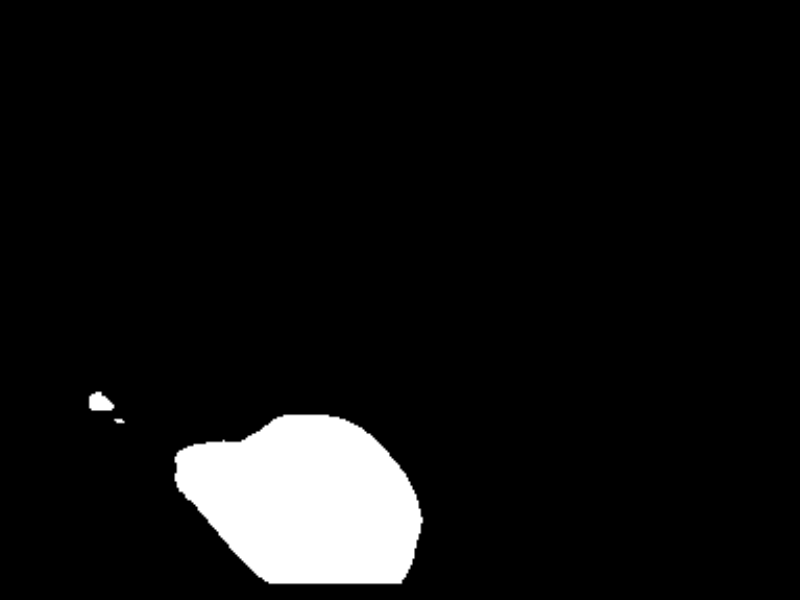}  
        \caption*{Attention-Unet}
    \end{subfigure}
    \begin{subfigure}[b]{0.15\textwidth}  
        \centering
        \includegraphics[width=0.8\textwidth]{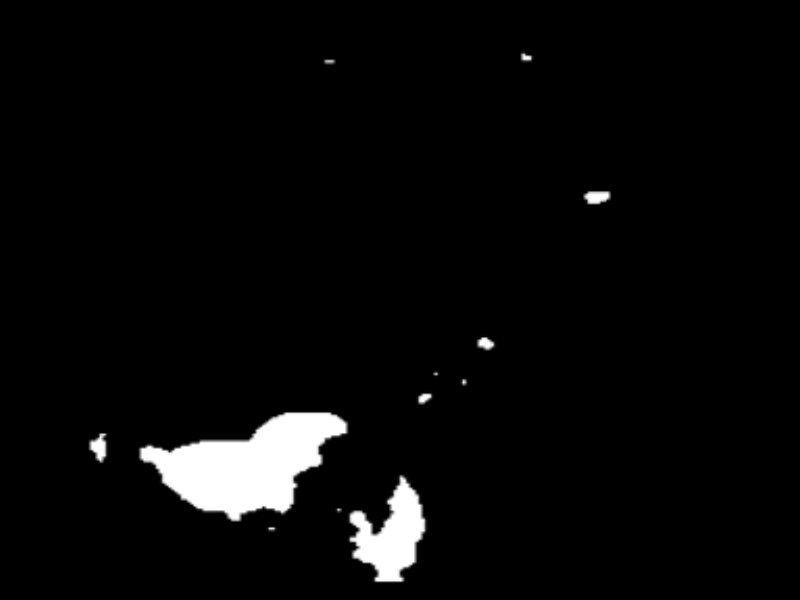}  
        \caption*{KiU-Net}
    \end{subfigure}
    \begin{subfigure}[b]{0.15\textwidth}  
        \centering
        \includegraphics[width=0.8\textwidth]{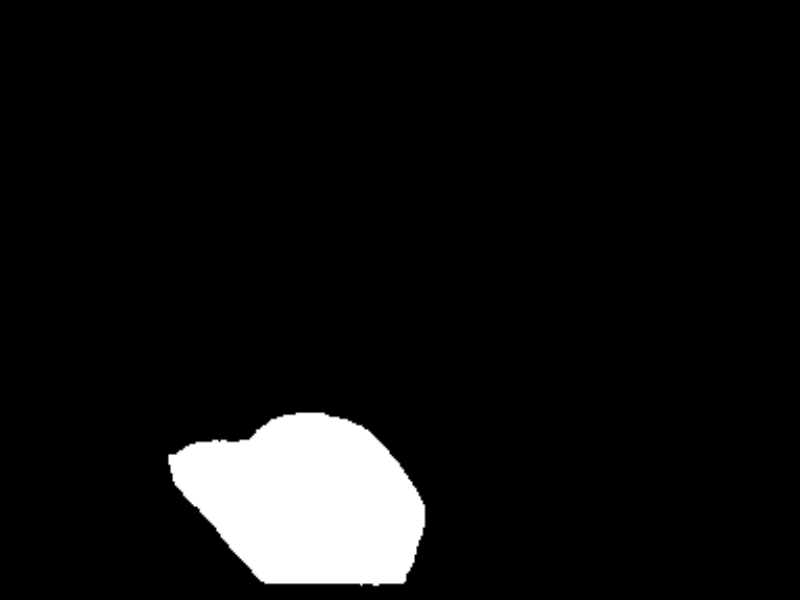}
        \caption*{MedFormer}
    \end{subfigure}
    \caption{Sample results of polyp segmentation. From left to right: input image, ground truth, SOTA results obtained by MGFI-Net, CE-Net, Attention U-Net, U-Net++, U-Net,
KiU-Net, MedFormer.}
    \label{fig5}
\end{figure*}

\begin{table*}[t]
\centering
\caption{Performance of polyp segmentation. The best and second-best results for each metric are highlighted in \textcolor{red}{red} and \textcolor{blue}{blue}, respectively. The ↑ symbol indicates that a higher metric score corresponds to better model performance.}
\begin{tabular}{lccccc}
	\toprule
    \multicolumn{6}{c}{CVC-ClinicDB}\\
    \midrule
		Method & Accuracy(↑) & Dice(↑)& IoU(↑)&Recall(↑)&Precision(↑) \\
    \midrule
    U-Net \cite{unet}          & 0.9791 & 0.8647 & 0.7957 &0.8813 &0.8835\\
	UNet++ \cite{u2+}         & 0.9802 & 0.8684 & 0.7977 &0.8485 &0.9201\\
	CE-Net \cite{ce}         & 0.9857 & 0.9233 & 0.8578 &0.9045 &\color{blue}0.9493\\
	Attention-Unet \cite{atunet} & 0.9821 & 0.8873 & 0.8215 &0.8898 &0.9055\\
	KiU-Net \cite{DBLP:journals/tmi/ValanarasuSHP22}        & 0.9770 & 0.8493 & 0.7546 &0.8527 &0.9225\\
	MedFormer \cite{utnet}      & \color{blue}0.9886 & \color{blue}0.9349 & \color{blue}0.8797 &\color{blue}0.9300 &0.9470\\
	MGFI-Net       & \color{red}0.9911 & \color{red}0.9497 & \color{red}0.9050 &\color{red}0.9463 &\color{red}0.9599\\
	\bottomrule
    \label{tab1}
\end{tabular}
\end{table*}

\subsection{Loss Function}
In medical image segmentation tasks, achieving precise segmentation, particularly at the boundaries of the target regions, is essential for high-performance models. To this end, we employ a hybrid loss function that combines Cross-Entropy Loss, Dice Loss\cite{dice}, and Boundary Loss. Each of these components plays a vital role in enhancing the overall segmentation accuracy, especially when dealing with complex medical images where both regional and boundary information are critical.

The hybrid loss function, $L_{hybrid}$, is formulated as \eqref{E1}:
\begin{equation}
    \label{E1}
   L_{\text{hybrid}} = L_{\text{Cross-entropy}} + L_{\text{Dice}} + \lambda \times L_{\text{Boundary}}
\end{equation}

Here, $L_{Cross-entropy}$ measures pixel-wise classification errors by comparing the predicted output with the ground truth labels. It is defined as \eqref{E2}:
\begin{equation}
    \label{E2}
   L_{\text{Cross-entropy}} = - \frac{1}{N} \sum_{c=0}^{C} \sum_{n=0}^{N} g_{n,c}^{\text{original}} \cdot \log(o_{n,c})
\end{equation}

Where \(N\) represents the total number of pixels, \(C\) is the number of classes, \(g_{n,c}^{\text{original}}\) is the ground truth label, and \(o_{n,c}\) is the predicted output for each pixel. This term ensures the ability of the model to correctly classify each pixel.

To handle the problem of class imbalance, especially when segmenting small target regions, Dice Loss is also employed. It is defined as \eqref{E3}:
\begin{equation}
    \label{E3}
L_{\text{Dice}} = 1 - \frac{2 \times \sum_{n=0}^{N} g_{n,c}^{\text{original}} \cdot o_{n,c}}{\sum_{n=0}^{N} g_{n,c}^{\text{original}} + \sum_{n=0}^{N} o_{n,c}}
\end{equation}
Finally, we introduce a Boundary Loss to further refine the prediction of target boundaries. This loss compares the predicted boundary with the ground truth boundary and is defined as \eqref{E4}:
\begin{equation}
\label{E4}
L_{\text{Boundary}} = \sum_{i=0}^{d} \left(1 - \sum_{c=0}^{C} \frac{2 \times \sum_{n=0}^{N} g_{n,c}^{\text{boundary}} \cdot o'_{n,c}}{\sum_{n=0}^{N} g_{n,c}^{\text{boundary}} + o'_{n,c}}\right)
\end{equation}

Here, $g_{n,c}^{boundary}$ represents the ground truth boundary labels, which are obtained using the Canny\cite{canny1986computational} edge detection operator, ensuring accurate delineation of the true boundary. On the other hand, $o'_{n,c}$ represents the predicted boundary output from the model. The weight coefficient $\lambda$ balances the contribution of the boundary loss in the overall hybrid loss function, ensuring the model focuses on both accurate regional segmentation and boundary precision.

By combining these loss components, the hybrid loss function provides a balanced supervision mechanism that not only ensures accurate pixel classification but also enhances boundary preservation. This approach leads to more precise segmentation results, making the model particularly effective in medical imaging scenarios with complex structures and irregular boundaries.

\section{Experiments}
\subsection{Datasets}
To fairly evaluate the proposed model, we used three public medical image segmentation datasets: CVC-ClinicDB\cite{polyp}, 2018 Data Science Bowl\cite{bowl}, and the International Skin Imaging Collaboration (ISIC) 2018 Challenge\cite{isic}\cite{Tschandl2018}. These datasets contain various challenging medical image segmentation tasks, such as colorectal polyp segmentation, nuclear segmentation in microscopy images, and dermoscopic skin lesion segmentation. The images in these datasets come from different resolutions, lighting conditions, and imaging angles, providing a rich diversity and challenge for segmentation tasks. 

\subsection{Baseline Models}
To validate the effectiveness of the proposed model on medical image datasets, we compared it with several state-of-the-art models. These models include U-Net, UNet++, Attention U-Net, CE-Net, KIU-Net, MedFormer. Note that these models are popular segmentation model for medical images.

\subsection{Evaluation Metrics}
In this study, we evaluate the segmentation performance of the model using common metrics: Precision, Recall, Accuracy, Dice, and IoU. Dice and IoU are particularly useful for measuring the overlap between predicted and ground truth masks, with Dice focusing on similarity and IoU quantifying the ratio of overlap to the union of the two areas. Accuracy measures the proportion of correct predictions, while Dice, also known as the F1 score, is effective in handling class imbalances by considering both true positive and false negative predictions.

The values for these evaluation metrics, including Accuracy, Dice, and IoU, range from 0 to 1, where 1 indicates a perfect match and 0 represents no match. Higher values indicate better segmentation performance. We use Floating Point Operations (FLOPs), Parameters (Params), and Frames Per Second (FPS) as metrics to assess the efficiency.

\subsection{Experimental Setup and Details}
All experiments were conducted on NVIDIA L20 GPU with 48GB memory, running Ubuntu 24.04.1 LTS, using PyTorch 1.13.0 framework for model training and evaluation. The encoder, based on a ResNet 34\cite{resnet} architecture, was used to extract multi-scale features across all models. Due to the varying sample sizes in the ISIC-2018 challenge datasets, we resized the images to 256x256. All datasets were split into 80\% for training, 10\% for validation, and 10\% for testing.

During training, we applied common data augmentation techniques to the training set to prevent overfitting. These techniques included center cropping, random rotation, transposition, and adding Gaussian noise. For the 2018 Data Science Bowl dataset and ISIC-2018, we set the batch size to 32 and the learning rate to 0.001. For datasets with larger image sizes (CVC-ClinicDB), we set the batch size to 8 and the learning rate to 0.0001. All experiments were conducted over 50 training epochs, using the Adam optimizer, with early stopping employed to prevent overfitting.

\begin{figure*}[t]
    \centering
    \begin{subfigure}[b]{0.15\textwidth}  
        \centering
        \includegraphics[width=0.8\textwidth]{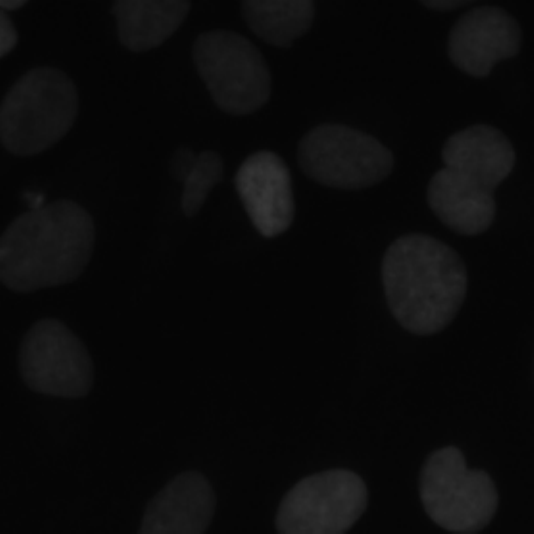} 
        \caption*{Input}
    \end{subfigure}
    \begin{subfigure}[b]{0.15\textwidth}  
        \centering
        \includegraphics[width=0.8\textwidth]{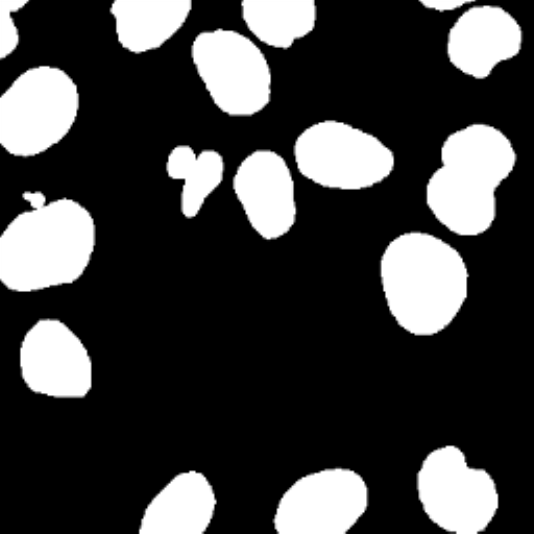} 
        \caption*{Ground Truth}
    \end{subfigure}
    \begin{subfigure}[b]{0.15\textwidth}  
        \centering
        \includegraphics[width=0.8\textwidth]{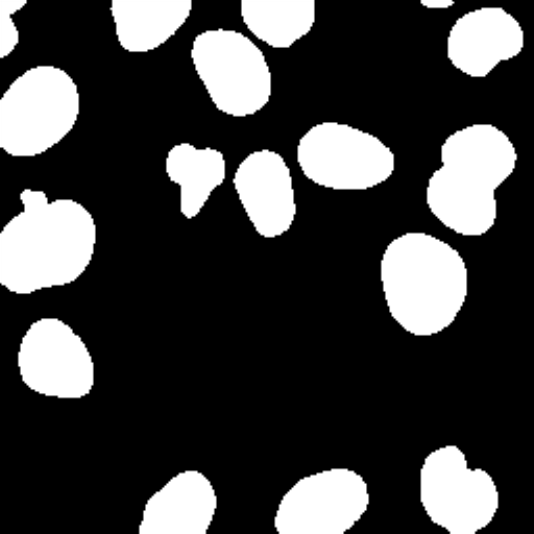}  
        \caption*{MGFI-Net}
    \end{subfigure}
    \begin{subfigure}[b]{0.15\textwidth}  
        \centering
        \includegraphics[width=0.8\textwidth]{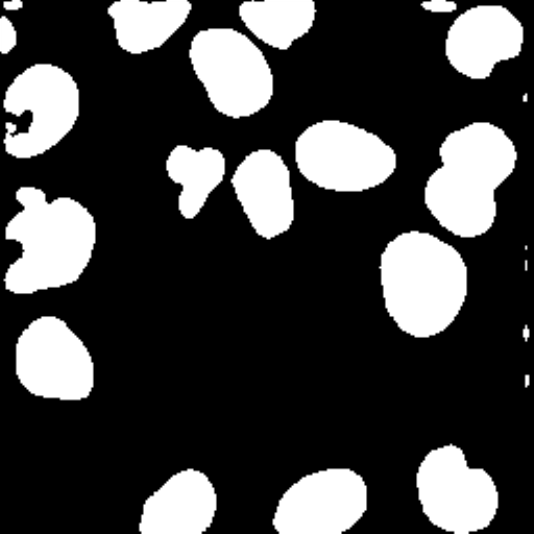}  
        \caption*{U-Net}
    \end{subfigure}
    \begin{subfigure}[b]{0.15\textwidth} 
        \centering
        \includegraphics[width=0.8\textwidth]{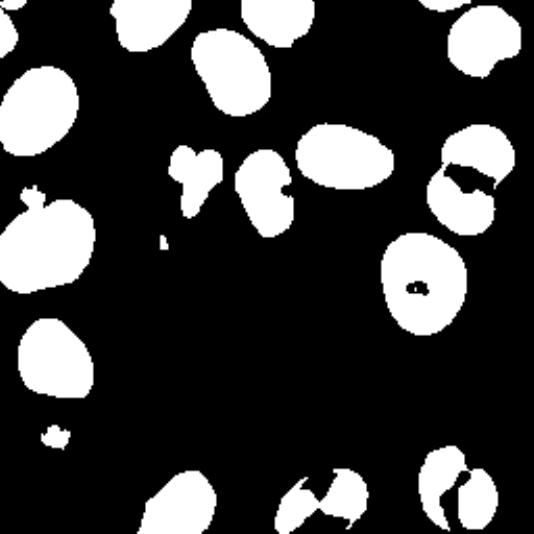}  
        \caption*{UNet++}
    \end{subfigure}
    \begin{subfigure}[b]{0.15\textwidth}  
        \centering
        \includegraphics[width=0.8\textwidth]{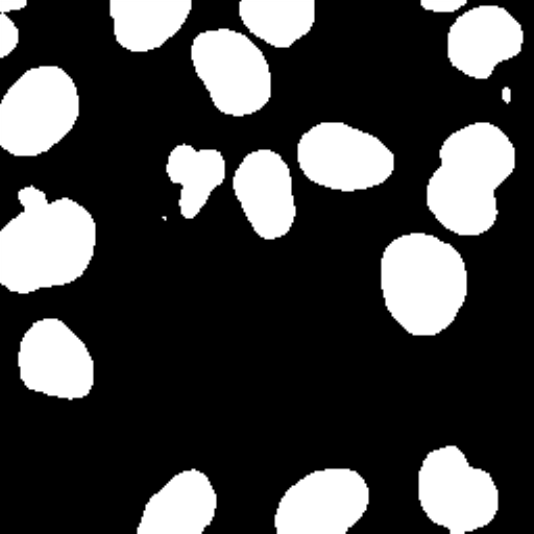}  
        \caption*{CE-Net}
    \end{subfigure}
    \begin{subfigure}[b]{0.15\textwidth} 
        \centering
        \includegraphics[width=0.8\textwidth]{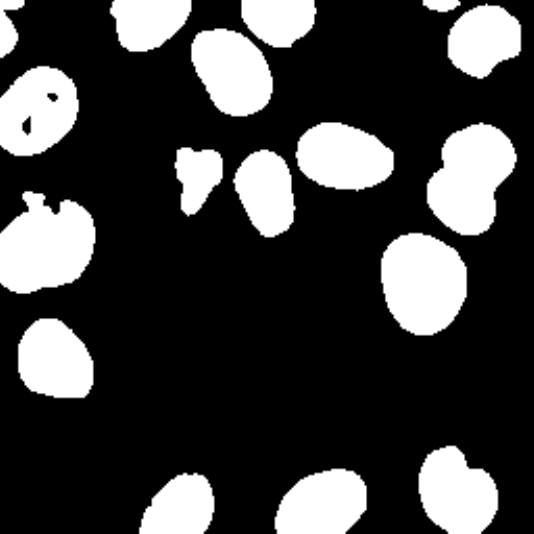}  
        \caption*{Attention-Unet}
    \end{subfigure}
    \begin{subfigure}[b]{0.15\textwidth}  
        \centering
        \includegraphics[width=0.8\textwidth]{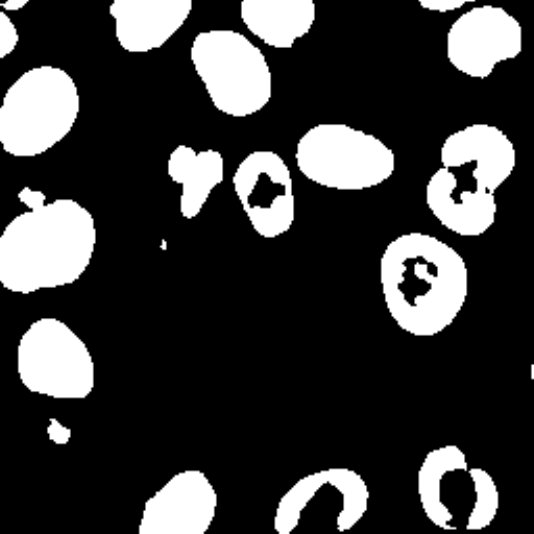}  
        \caption*{KiU-Net}
    \end{subfigure}
    \begin{subfigure}[b]{0.15\textwidth}  
        \centering
        \includegraphics[width=0.8\textwidth]{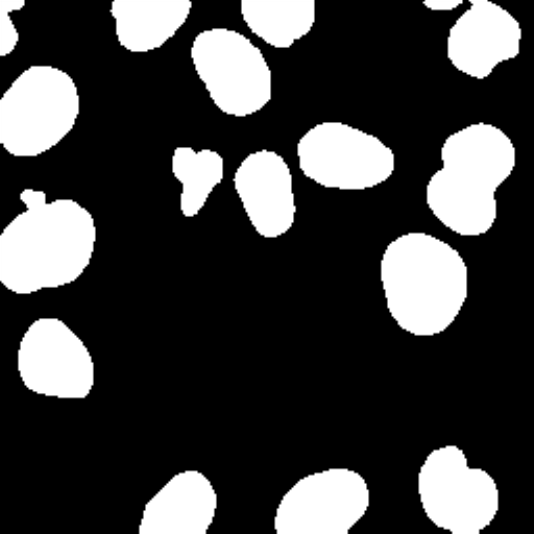}
        \caption*{MedFormer}
    \end{subfigure}
    \caption{Sample results of nuclei segmentation. From left to right: input image, ground truth, SOTA results obtained by MGFI-Net, CE-Net, Attention U-Net, U-Net++, U-Net,
KiU-Net, MedFormer.}
    \label{fig6}
\end{figure*}
\begin{table*}[t]
\centering
\caption{Performance of nuclei segmentation. The best and second-best results for each metric are highlighted in \textcolor{red}{red} and \textcolor{blue}{blue}, respectively. The ↑ symbol indicates that a higher metric score corresponds to better model performance.}
\begin{tabular}{lccccc}
	\toprule
    \multicolumn{6}{c}{2018 Data Science Bowl}\\
    \midrule
		Method & Accuracy(↑) & Dice(↑)& IoU(↑)&Recall(↑)&Precision(↑) \\
    \midrule
    U-Net \cite{unet}         & 0.9739 & 0.8901 & 0.8136 &0.8996 &0.8896\\
	UNet++ \cite{u2+}        & 0.9729 & 0.8894 & 0.8197 &0.8976 &0.8914\\
	CE-Net \cite{ce}        & 0.9751 & 0.9133 & 0.8423 &0.9153 &0.9048\\
	Attention-Unet \cite{atunet} & 0.9753 & 0.8905 & 0.8220 &0.8969 &0.8879\\
	KiU-Net \cite{DBLP:journals/tmi/ValanarasuSHP22}       & 0.9789 & 0.8419 & 0.4912 &\color{red}0.9478 &0.8980\\
	MedFormer \cite{atunet}     & \color{blue}0.9815 & \color{blue}0.9134 & \color{blue}0.8512 &0.9145 &\color{blue}0.9275\\
	MGFI-Net       & \color{red}0.9817 & \color{red}0.9206 & \color{red}0.8571 &\color{blue}0.9250 &\color{red}0.9493\\
	\bottomrule
    \label{tab2}
\end{tabular}
\end{table*}

\begin{figure*}[t]
    \centering
    \begin{subfigure}[b]{0.15\textwidth}  
        \centering
        \includegraphics[width=0.8\textwidth]{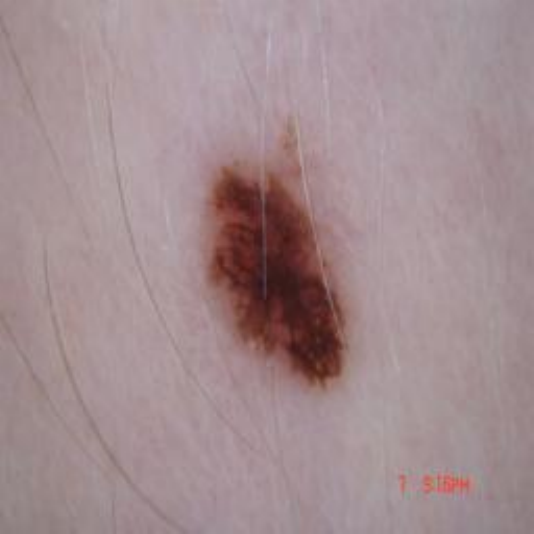} 
        \caption*{Input}
    \end{subfigure}
    \begin{subfigure}[b]{0.15\textwidth}  
        \centering
        \includegraphics[width=0.8\textwidth]{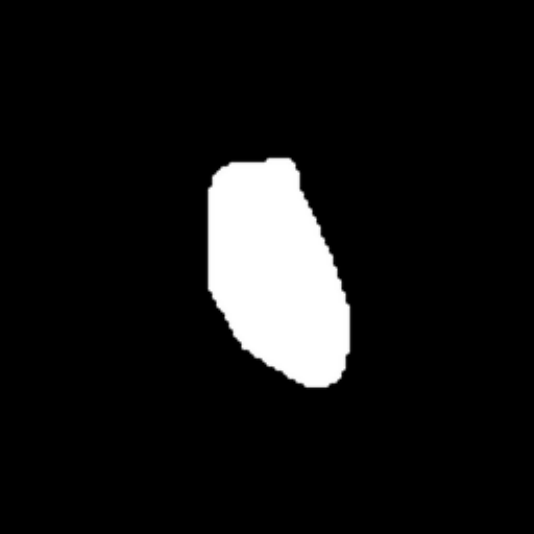} 
        \caption*{Ground Truth}
    \end{subfigure}
    \begin{subfigure}[b]{0.15\textwidth}  
        \centering
        \includegraphics[width=0.8\textwidth]{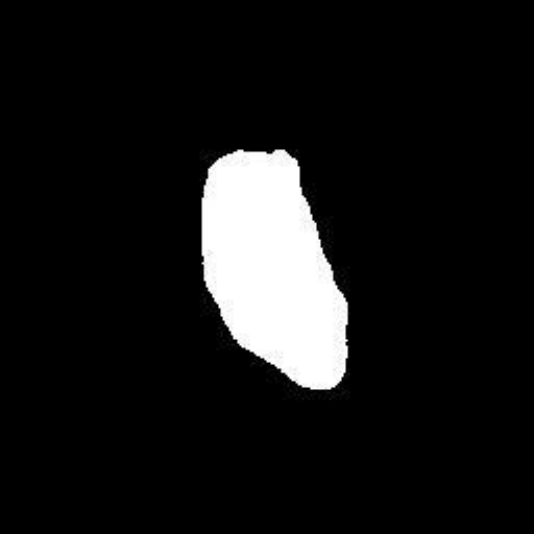}  
        \caption*{MGFI-Net}
    \end{subfigure}
    \begin{subfigure}[b]{0.15\textwidth}  
        \centering
        \includegraphics[width=0.8\textwidth]{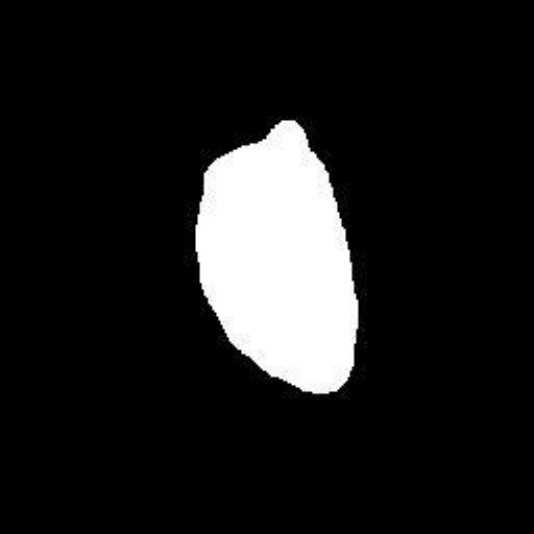}  
        \caption*{U-Net}
    \end{subfigure}
    \begin{subfigure}[b]{0.15\textwidth} 
        \centering
        \includegraphics[width=0.8\textwidth]{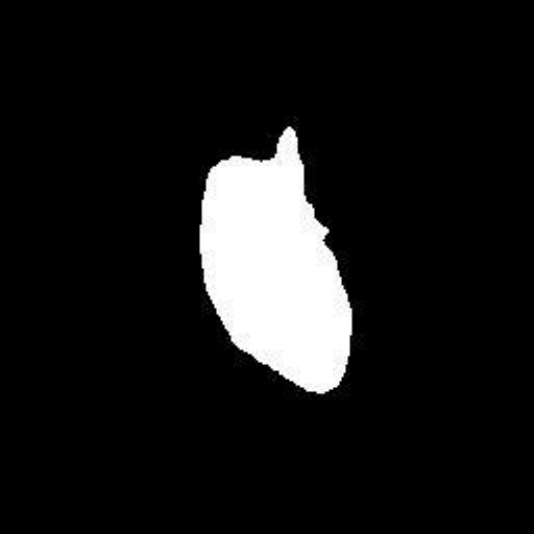}  
        \caption*{UNet++}
    \end{subfigure}
    \begin{subfigure}[b]{0.15\textwidth}  
        \centering
        \includegraphics[width=0.8\textwidth]{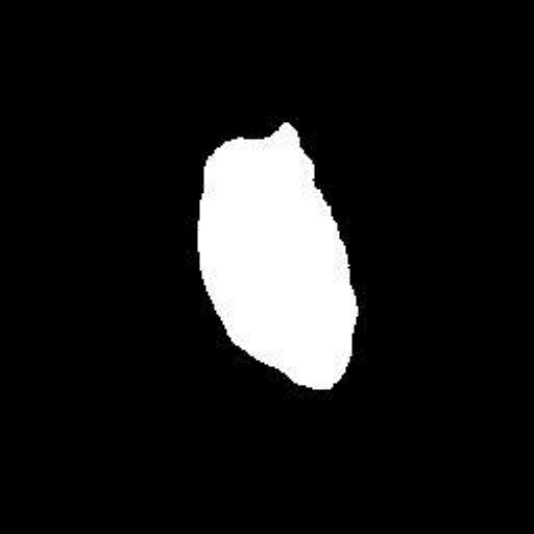}  
        \caption*{CE-Net}
    \end{subfigure}
    \begin{subfigure}[b]{0.15\textwidth} 
        \centering
        \includegraphics[width=0.8\textwidth]{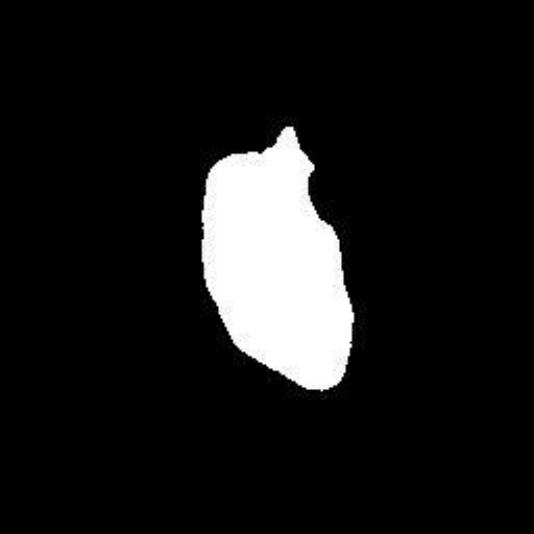}  
        \caption*{Attention-Unet}
    \end{subfigure}
    \begin{subfigure}[b]{0.15\textwidth}  
        \centering
        \includegraphics[width=0.8\textwidth]{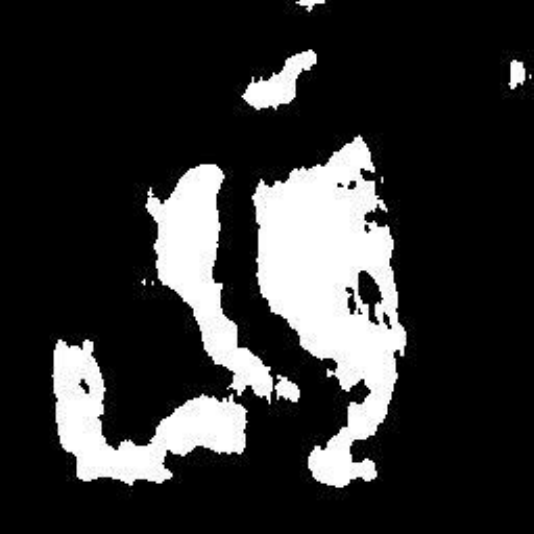}  
        \caption*{KiU-Net}
    \end{subfigure}
    \begin{subfigure}[b]{0.15\textwidth}  
        \centering
        \includegraphics[width=0.8\textwidth]{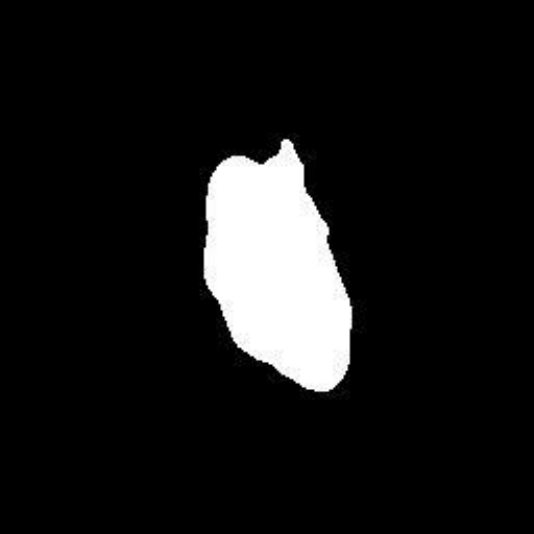}  
        \caption*{MedFormer}
    \end{subfigure}
    \caption{Sample results of skin lesion segmentation. From left to right: input image, ground truth, SOTA results obtained by MGFI-Net, CE-Net, Attention U-Net, U-Net++, U-Net,
KiU-Net, MedFormer.}
    \label{fig7}
\end{figure*}

Except for the KiU-Net model, all other models were trained under the same configuration. Due to the high memory requirement of KiU-Net, We followed the configuration of the original paper for this model.

\subsection{Comparisons with State-of-the-Art Methods}
In this section, we conduct a comparative analysis of our model against six state-of-the-art segmentation models across three widely-used public datasets. The comparative results and analysis are shown as follows.

\textbf{Results on CVC-ClinicDB}. Our MGFI-Net outperforms other segmentation models on the CVC-ClinicDB dataset, as shown in Fig.~\ref{fig5} and Table \ref{tab1}. MGFI-Net accurately segments the polyp region without background noise, especially in the elongated lower-left section, unlike models like UNet++, CE-Net, and Attention-Unet, which include background as part of the polyp. 

While CE-Net excels in accuracy and precision, it struggles with fine details, particularly at the polyp's tip. MedFormer performs well but fails to capture intricate boundary details. MGFI-Net provides the closest segmentation to the ground truth, handling challenging regions and maintaining boundary precision, demonstrating its robustness in complex medical images.

\textbf{Results on 2018 Data Science Bowl}. Our MGFI-Net outperforms other state-of-the-art segmentation models on the 2018 Data Science Bowl dataset, as shown in the visual comparison in Fig.~\ref{fig6} and the quantitative results in Table \ref{tab2}. Notably, MGFI-Net accurately segments the nuclei without including noise or merging cells, especially in areas with closely packed or faintly visible nuclei.

CE-Net introduces noise, leading to false positives, as seen in Fig.~\ref{fig6}, where scattered noise pixels are misclassified as nuclei. U-Net, UNet++, and Attention-Unet also face challenges, misclassifying some nuclei as hollow structures, particularly in regions with indistinct boundaries or clustered cells. These issues impact segmentation accuracy.

\begin{table*}[t]
\centering
\caption{Performance of skin lesion segmentation. The best and second-best results for each metric are highlighted in \textcolor{red}{red} and \textcolor{blue}{blue}, respectively. The ↑ symbol indicates that a higher metric score corresponds to better model performance.}
\begin{tabular}{lccccc}
	\toprule
    \multicolumn{6}{c}{ISIC-2018 challenge}\\
    \midrule
		Method & Accuracy(↑) & Dice(↑)& IoU(↑)&Recall(↑)&Precision(↑) \\
    \midrule
    U-Net \cite{unet}& 0.9009 & 0.8483 &  0.8647 &\color{blue}0.8847 &0.8395\\
	UNet++ \cite{u2+}& 0.9041 & 0.8568 & 0.7047 &0.8738 &0.8547\\
	CE-Net \cite{ce}        & \color{blue}0.9194 & \color{blue}0.8745 & \color{blue}0.7171 &0.8846 &\color{blue}0.8836\\
	Attention-Unet \cite{atunet}& 0.9052 & 0.8568 & 0.7089 &0.8697 &0.8694\\
	KiU-Net \cite{DBLP:journals/tmi/ValanarasuSHP22}       & 0.7995 & 0.6806 & 0.4912 &0.8301 &0.6473\\
	MedFormer \cite{utnet}     & 0.9183 & 0.8712 & 0.7129 &0.8652 &0.8679\\
	MGFI-Net       & \color{red}0.9384 & \color{red}0.8988 & \color{red}0.7734 &\color{red}0.9134 &\color{red}0.8956\\
	\bottomrule
    \label{tab3}
\end{tabular}
\end{table*}

Overall, as illustrated in Fig.~\ref{fig6}, MGFI-Net’s segmentation results closely match the ground truth, outperforming other models in maintaining both completeness and precision. Its robustness in handling noisy environments and complex cell structures ensures more accurate and reliable segmentation, as further supported by the quantitative results in Table \ref{tab2}.

\begin{table*}[t]
\centering
\caption{Comparison on CVC-ClinicDB. The comparison metrics are the FLOPs, Params, Dice and FPS. The best and second-best results for each metric are highlighted in red and blue, respectively. The ↑ symbol indicates that a higher metric score corresponds to better model performance. The ↓ symbol indicates that a lower metric score corresponds to better model performance.}
\begin{tabular}{lllcl}
	\toprule
Method & FLOPs(↓) & Params(↓)& Dice(↑)&FPS(↑) \\
    \midrule
U-Net \cite{unet}& \color{red}12155.35 & 7.77 & 0.8647 & \color{red}184.51\\
UNet++ \cite{u2+}& 33928.50 & 9.18 & 0.8684 &58.72 \\
CE-Net \cite{ce} & \color{blue}16367.31 & 10.23 &  0.9233 &54.76 \\
Attention-Unet \cite{atunet}& 67363.16 & 34.93 & 0.8873 &38.65 \\
KiU-Net \cite{DBLP:journals/tmi/ValanarasuSHP22}& 517363.04 & \color{red}0.29 & 0.8493 &3.97\\
MedFormer \cite{utnet}& 39427.40 & 9.73 & \color{blue}0.9349  &4.11 \\
MGFI-Net  & 21589.15 & \color{blue}7.65 & \color{red}0.9497 & \color{blue}101.45\\
	\bottomrule
    \label{tab33}
\end{tabular}
\end{table*}

\begin{table*}[t]
\centering
\caption{Performance of ablation study on the MGFI module. The best and second-best results for each metric are highlighted
in red and blue, respectively. The ↑ symbol indicates that a higher metric score corresponds to better model performance.}
\begin{tabular}{lccccc}
	\toprule
		Method & Accuracy(↑) & Dice(↑)& IoU(↑)&Recall(↑)&Precision(↑) \\
    \midrule
        Variant 1 (Upper section of MGFI module) &0.9843&0.9055&0.8676&0.9017&0.9151\\
        Variant 2 (Lower section of MGFI module) 
        &\textcolor{blue}{0.9879}&\textcolor{blue}{0.9410}&\textcolor{blue}{0.8862}&\textcolor{blue}{0.9367}&\textcolor{blue}{0.9512}\\
        Variant 3 (MGFI module) &0.9828&0.8990&0.8645&0.8991&0.9062\\
        MGFI-Net                &\textcolor{red}{0.9911}&\textcolor{red}{0.9497}&\textcolor{red}{0.9050}&\textcolor{red}{0.9463}&\textcolor{red}{0.9599}\\
	\bottomrule
    \label{tab4}
\end{tabular}
\end{table*}
\begin{table*}[t]
\centering
\caption{Performance of ablation study on the AE module. The best and second-best results for each metric are highlighted in \textcolor{red}{red} and \textcolor{blue}{blue}, respectively. The ↑ symbol indicates that a higher metric score corresponds to better model performance.}
\begin{tabular}{lccccc}
	\toprule
		Method & Accuracy(↑) & Dice(↑)& IoU(↑)&Recall(↑)&Precision(↑) \\
    \midrule
        Variant 1 (AE module) &\textcolor{blue}{0.9864}&\textcolor{blue}{0.9219}&\textcolor{blue}{0.8589}&\textcolor{blue}{0.9309}&\textcolor{blue}{0.9188}\\
        MGFI-Net              &\textcolor{red}{0.9911}&\textcolor{red}{0.9497}&\textcolor{red}{0.9050}&\textcolor{red}{0.9463}&\textcolor{red}{0.9599}\\
	\bottomrule
    \label{tab5}
\end{tabular}
\end{table*}

\textbf{Results on ISIC-2018 challenge}. MGFI-Net outperforms other segmentation models on the ISIC-2018 challenge dataset, as shown in Fig.~\ref{fig7} and Table \ref{tab3}. MGFI-Net accurately segments the lesion, avoiding misidentifying noise as part of the target, unlike models such as U-Net, UNet++, and CE-Net.

MGFI-Net excels with multi-grained features and precise edge detection, reducing false positives. CE-Net, despite high accuracy, struggles with recall, missing smaller regions. MedFormer lacks fine detail, lowering Dice and IoU scores, while KiU-Net fails to balance local and broader features, leading to poor segmentation.

Overall, MGFI-Net demonstrates superior performance in accurate lesion segmentation, effectively capturing lesion boundaries and avoiding false positives.

\subsection{Efficiency analysis on CVC-ClinicDB dataset}  
To assess the efficiency of Image Semantic Segmentation (ISS) models, we compare MGFI-Net with baseline models in terms of FLOPs, Params, Dice, and FPS. As shown in Table \ref{tab33}, MGFI-Net achieves the highest Dice score of 94.97\% and an impressive FPS of 101.45, ranking second in inference speed.

Compared to U-Net, which has the highest FPS, MGFI-Net shows an 8.5\% improvement in Dice. While U-Net achieves faster inference, it has lower accuracy, indicating limitations in complex tasks.

Compared to MedFormer, MGFI-Net demonstrates better efficiency, with similar Dice but significantly fewer FLOPs and Params, showing global context awareness with lower computational costs. MedFormer’s high FLOPs (39,427.40) and low FPS (4.11) reveal inefficiency.

KiU-Net, despite having the smallest parameter count (0.29M), has high FLOPs (517,363.04) and low FPS (3.97), with a Dice score of 84.93\%, which is lower than MGFI-Net, indicating poor segmentation accuracy and inefficiency.

In summary, MGFI-Net offers a better balance of accuracy, efficiency, and computational cost, making it a promising solution for ISS tasks in medical images. Further optimization could enhance its real-time performance.

\subsection{Ablation study on CVC-ClinicDB dataset}
To validate the contribution of each component of the proposed MGFI-Net, we conduct ablation experiments on the CVC-ClinicDB dataset. The goal of these experiments is to demonstrate the effectiveness of both the Multi-Grained Feature Integration (MGFI) module and the Adaptive Edge (AE) module, which are critical for achieving high segmentation accuracy.

The ablation study on the MGFI module (as shown in Table \ref{tab4}) highlights the impact of different sections of the module on overall performance. When the upper part of the MGFI module, which handles overlapping downsampling and multi-branch feature extraction, is removed, the performance of the model declines. This decline is particularly noticeable in Dice and IoU scores. These results indicate the importance of this section in preserving spatial continuity and integrating features from various regions. Similarly, removing the lower part significantly drops performance. This section refines features using deformable, atrous, and standard convolutions. The decrease is particularly evident in Dice and Recall metrics. This suggests that multi-scale refinement is crucial for capturing complex shapes and granular details accurately. The most substantial degradation occurs when the entire MGFI module is removed, confirming that the integration of multi-grained features plays a critical role in capturing both spatial details and contextual relationships. The full MGFI-Net, with both sections intact, achieves the highest scores across all metrics, demonstrating the effectiveness of combining these components for accurate and robust segmentation of medical images.

The ablation study on the AE (Adaptive Edge) module, as presented in Table \ref{tab5}, demonstrates the importance of edge refinement in improving segmentation performance. When the AE module is removed (Variant 1), the model experiences a significant drop in Dice and IoU scores, particularly with a Dice score decrease from 0.9497 to 0.9219 and an IoU decrease from 0.9050 to 0.8589. This reduction highlights the critical role the AE module plays in accurately preserving edge information, especially in challenging regions with irregular boundaries. Although the model without the AE module still achieves relatively high Recall, the lower Precision indicates that the absence of dynamic edge refinement leads to more false positives, reducing the overall segmentation accuracy. By incorporating deformable convolutions, the AE module dynamically refines boundary information, allowing MGFI-Net to maintain high Precision (0.9599) and accurately capture complex edge structures, leading to superior performance in medical image segmentation.

The results of the ablation experiments demonstrate that the proposed MGFI module and AE module are essential for improving segmentation performance in complex medical imaging tasks. The MGFI module effectively captures multi-grained features, enhancing the ability of the model to retain fine spatial details and broader contextual information. Meanwhile, the AE module significantly improves the capacity of the model to handle irregular boundaries and complex edge structures by refining boundary details. The combination of these two modules leads to the highest Accuracy, Dice, and IoU scores, as shown in the experiments on the CVC-ClinicDB dataset.
\section{Conclusion}
In this paper, we propose MGFI-Net, a model that improves medical image segmentation by leveraging multi-grained information and adaptively learning edge features. By capturing features at different levels of granularity, the model is able to accurately focus on the most relevant information for segmentation, especially in complex images with intricate structures or noise. Additionally, MGFI-Net enhances the preservation of fine edge details, which are often blurred or lost in traditional models. Extensive experiments conducted on three public medical image datasets demonstrate that MGFI-Net consistently outperforms state-of-the-art segmentation models. These results confirm the effectiveness of MGFI-Net, making it highly applicable to a wide range of clinical applications.

In future work, we plan to extend the application of our model by conducting research in MRI and CT image segmentation.

{\small
\bibliographystyle{cvm}
\bibliography{cvmbib_my}
}

\end{document}